\documentclass[twocolumn]{autart}    

\usepackage{graphicx}          
\usepackage{amssymb}
\usepackage{amsmath}
\usepackage{bm}
\usepackage{stfloats}
\usepackage{graphicx}
\usepackage[ruled]{algorithm}
\usepackage{epstopdf}
\usepackage{textcomp}
\usepackage{subcaption}
\usepackage{enumerate}
\usepackage{subeqnarray}

\usepackage{xcolor,calc}

\newcommand{\qedsymbol}{\hfill$\blacksquare$}

\newtheorem{myDef}{Definition}

\newtheorem{standing}{Standing Assumption}
\newtheorem{lemma}{Lemma}
\newtheorem{remark}{Remark}

\begin{document}

\begin{frontmatter}

\title{Identification of non-causal systems\\ with arbitrary switching modes} 


\author[Bit]{Yanxin Zhang}\ead{zhangyanxin@bit.edu.cn},   \author[Bit]{Chengpu Yu}\ead{yuchengpu@bit.edu.cn}, and \author[MIT]{Filippo Fabiani}\ead{filippo.fabiani@imtlucca.it}

\address[Bit]{School of Automation, Beijing Institute of Technology, Beijing 100081, PR China}  
\address[MIT]{IMT School for Advanced Studies Lucca, Piazza San Francesco 19, 55100, Lucca, Italy}                                           
\thanks{This work was supported by the National Natural Science Foundation of China (Grant No. 61991414, 62088101, 6193000461), Chongqing Natural Science Foundation CSTB\\2023NSCQ-JQX0018, and Beijing Natural Science Foundation L221005. Corresponding author: Chengpu Yu.}

\begin{abstract}                          
	
	We consider the identification of non-causal systems with arbitrary switching modes (NCS-ASM), a class of models essential for describing typical power load management and department store inventory dynamics. The simultaneous identification of causal-and-anticausal subsystems, along with the presence of possibly random switching sequences, however, make the overall identification problem particularly challenging. To this end, we develop an expectation-maximization (EM) based system identification technique, where the E-step proposes a modified Kalman filter (KF) to estimate the states and switching sequences of causal-and-anticausal subsystems, while the M-step consists in a switching least-squares algorithm to estimate the parameters of individual subsystems. We establish the main convergence features of the proposed identification procedure, also providing bounds on the parameter estimation errors under mild conditions. Finally, the effectiveness of our identification method is validated through two numerical simulations.
\end{abstract}

\end{frontmatter}

\section{Introduction}
Non-causal switching dynamics denote a class of systems able to modeling several real-world scenarios, such as load management in power systems \cite{Tan2023}, traffic signal control systems \cite{Liao2024}, and robotic systems \cite{Carloni2007,Schlegl2003}. These systems encompass non-causality, meaning that their output signals depend not only on the current or past control actions, but also on future inputs. In addition, the systems exhibit switching characteristics, potentially transitioning among different operational states, and thus leading to variations in the system behavior. Understanding and managing the complexity of these systems is therefore crucial for enhancing efficiency, reliability, and adaptability, enabling them to better meet the demands of industrial production and operations. This essentially motivates the interest in modeling, analyzing, and controlling such type of systems.

In many parameter identification problems for dynamical systems, the input-output data are accompanied by temporal mode sequences. As the mode of the system changes over time, each data point corresponds to the mode at its corresponding time instant. In these circumstances, it is hence crucial to model the dynamics of different modes and estimate the transitions from one mode to another \cite{Chan2008}. However, obtaining direct estimates of the dynamical system from input-output data is challenging and, in practice, the prior knowledge on the mode transitions is often unavailable. Therefore, estimating the switching behaviors poses a challenging, yet highly significant, problem addressed by several researchers. Available works indeed propose algorithms to estimate the individual system dynamics and the mode transition sequence based on observed behaviors \cite{Ferrari2003}.

\subsection{Literature review}
Several works consider the identification of switching models \cite{Garulli2012,Bianchi2021}. In \cite{Mark2022}, a joint smoothing algorithm is proposed based on the expectation-maximization (EM) framework, where an E-step solution is introduced to effectively address issues related to the exponential complexity in the jump Markov linear model. In \cite{Bemporad2018}, a numerically efficient, two-step estimation method was proposed, which iteratively updates the parameters and the switching sequence. The flexibility of this technique consists in its adaptability to different loss functions employed in the jump models, which significantly impact the overall shape and jumping behavior of the models. Furthermore, the identification of jump Box-Jenkins systems is investigated in \cite{Piga2020}, where a maximum a-posteriori method is proposed to estimate the switching sequence of the model. Subsequently, the system parameters of the jump Box-Jenkins models are alternately estimated using the Gauss-Newton and the prediction error methods. In \cite{Sayedana2024}, a switching least-squares algorithm for autonomous Markov jump linear systems is proposed. Here, the authors provided a formal proof characterizing the strong consistency of the underlying method, as well as established its convergence rate as $\mathcal{O}(\sqrt{\log(T)/T})$ a.s., where $T$ is the time horizon. The aforementioned literature primarily focuses on linear systems and assumes that the switching behavior follows a Markov chain. However, these techniques are not applicable when the mode switching is \textit{random}. Consequently, a number of approaches for identifying systems with random mode switching behavior have been proposed in the literature. For example, in \cite{Anna2018} a kernel-based method is used to estimate the random switching system, which can solve both estimation and classification problems simultaneously. In \cite{Angelo2010}, a maximum-likelihood algorithm is presented for the identification of a random switching linear system, which combines the maximum-likelihood estimation criterion and the Kalman filtering technique to estimate the system modes in switched linear systems, resulting in a significant contribution to the estimation error stability of general switched linear systems.

All the studies mentioned above are focused on switching causal systems, where the system output is only related to current or past excitation. However, to the best of our knowledge, there is no literature on the system identification problem of non-causal systems with arbitrary switching modes (NCS-ASM). Nevertheless, NCS-ASM are widely present in real-world scenarios. For instance, in automatic control systems the time delay caused by sensors can lead to the switching non-causal characteristics of the system \cite{Tsurumoto2022}. In financial markets, there exist interaction and feedback between investor decisions and market prices \cite{El2023}. Although, there are some studies available on system identification for non-causal systems, such as the subspace \cite{Verhaegen1996} and the kernel methods \cite{Fang2024,Blanken2020}, these studies can only handle a single, non-causal system, rather than \textit{switching non-causal systems}.

\subsection{Summary of contribution}

In this paper, we focus on the identification of NCS-ASM. The proposed method is developed under the expectation-maximization (EM) framework, which can be divided into two main parts. Specifically, in the E-step we adopt a Bayesian rule to compute the posterior estimate of the switching sequence, along with a modified Kalman filter (KF) for estimating the state of the causal and anti-causal parts. In the M-step, instead, we propose a switching least-squares method to obtain the closed-form solution for the parameters and establish the convergence rate of the estimated parameters. Our main contributions can hence be summarized as follows:
\begin{enumerate}
	\item To the best of our knowledge, this is the first work considering the idenfitication of NCS-ASM. In particular, in the E-step of the EM framework a modified KF is proposed to compute the posterior state estimates of the causal and anti-causal parts, which is crucial to deal with the identification of non-causal systems;
	\item Compared with the system dynamics in \cite{Mark2022,Bemporad2018,Piga2020}, where the switching behavior of the subsystems only occurs in the causal part, our methodology can cope with switching behaviors in both causal and anti-causal parts. Moreover, the switching sequences of the two directions is allowed to differ from each other. 
\end{enumerate}

\subsection{Paper organization}
The rest of the paper is organized as follows:
in Section \ref{sec3} we describe the considered system and formulate the related identification problem. In Section \ref{sec4}, instead, we discuss our EM method for the identification of the NCS-ASM, while in Section \ref{sec5} we provide its implementation details, as well as characterize the related convergence properties. Two simulation examples are finally discussed in Section \ref{sec6} to test the effectiveness of the proposed method numerically. 
The proofs of the technical results of the paper are all deferred to Appendix~\ref{sec:proofs}.

\textit{Notations:}
$\mathbb{Z}$ and $\mathbb{R}$ denote the set of integer and real numbers, respectively. 
Given a matrix $X$, $\Vert X\Vert$ and $\Vert X\Vert_{\infty}$ denote respectively its spectral and infinity norms, $\lambda_{\textrm{max}}(X)$ and $\lambda_{\textrm{min}}(X)$ are respectively its maximum and minimum eigenvalues, and $\textrm{tr}(X)$ denotes the trace. $\mathbb{P}[\cdot]$ and $\mathbb{E}[\cdot]$ denote respectively a distribution probability and the related expected value. $\mathbb{S}^{n}$ is the space of $n \times n$ symmetric matrices and $\mathbb{S}_{\succ 0}^{n}$ ($\mathbb{S}_{\succcurlyeq 0}^{n}$) is the cone of positive (semi-)definite matrices. Given two square matrices $A$, $B$ of compatible dimension, $A\succcurlyeq B$ means that $A-B$ is positive semidefinite. For a sequence $\{s_t\}_{t\in\mathbb{N}}$, $s_T=\mathcal{O}(T)$ indicates that $\lim\sup_{T\rightarrow\infty}s_T/T\textless\infty$, while $s_T=o(T)$ that $\lim\sup_{T\rightarrow\infty}s_T/T=0$. Finally, $I$ identifies a standard identity matrix. $\mathcal{N}(\mu,\sigma^2)$ denotes the normal distribution of a random variable with mean $\mu$ and standard deviation $\sigma$.
 
\section{Mathematical formulation}\label{sec3}
We now describe the system concerned in this paper, together with the main assumptions, and successively formalize the problem to be addressed.
\subsection{System model description}
Consider the following discrete-time, non-causal system characterized by arbitrary switching modes:
\begin{subeqnarray}\label{eq:1}
	x_c(t)&=&A_c(s_c(t))x_c(t-1)+v_c(t),\\
	x_a(t)&=&A_a(s_a(t))x_a(t+1)+v_a(t),\\
	y(t)&=&C_c(s_c(t))x_c(t)+C_a(s_a(t))x_a(t)+v_m(t),\nonumber\\
\end{subeqnarray}
where $t\in\mathbb{Z}$ is the time instant, $x_c(t)\in\mathbb{R}^{n_{x_c}},x_a(t)\in\mathbb{R}^{n_{x_a}}$ are the causal and anti-causal state vectors, respectively, $y(t)\in\mathbb{R}^{n_y}$ denotes the system output, while $s_c(t)\in\{1,2,\ldots,m_c\}\triangleq\Lambda_c$ and $s_a(t)\in\{1,2,\ldots,m_a\}\triangleq\Lambda_a$ are two discrete variables representing the possible switching modes. In addition, $v_c(t)\in\mathbb{R}^{n_{x_c}}$ and $v_a(t)\in\mathbb{R}^{n_{x_a}}$ are the system noise vectors, and $v_m(t)\in\mathbb{R}^{n_y}$ is the measurement noise vector. Finally, $A_c:\Lambda_c\to\mathbb{R}^{n_{x_c}\times n_{x_c}}$ and $A_a:\Lambda_a\to\mathbb{R}^{n_{x_a}\times n_{x_a}}$ denote the matrix functions associated to the causal and anti-causal state dynamics, respectively, while $C_c:\Lambda_c\to\mathbb{R}^{n_{y}\times n_{x_c}}$ and $C_a:\Lambda_a\to\mathbb{R}^{n_{y}\times n_{x_a}}$ are those mapping the two state vectors to the measured output. 

Assume that the noise terms $v_c(t)$, $v_a(t)$ and $v_m(t)$ are distributed according to a Gaussian distribution with zero mean and finite variance as follows:
\begin{align*}
	v_c(t)&\sim\mathcal{N}(0,\Sigma_c(s_c(t))),\\
	v_a(t)&\sim\mathcal{N}(0,\Sigma_a(s_a(t))),\\
	v_m(t)&\sim\mathcal{N}(0,\Sigma_m).
\end{align*}

\begin{standing}\label{assumpt2}
	The NCS-ASM \eqref{eq:1} is stable in the average sense, which means
	\[
		\sum_{i=1}^{T}\Vert x_c(i)\Vert^2=\mathcal{O}(T),\quad \sum_{i=1}^{T}\Vert x_a(i)\Vert^2=\mathcal{O}(T). 
	\]
	where $T$ is the sample size of the available dataset.
\end{standing}

\begin{remark}
	Stability in the average sense is widely applied in linear systems \cite{Sayedana2024,sta_as1,sta_as2}. Note that, compared to other commonly used notions, such as mean-square stability and almost sure stability, the assumption of stable in the average sense is weaker.
\end{remark}
\begin{standing}\label{assump1}
	\sloppy The noise vectors $v_c(t),v_a(t),v_m(t)$ are the martingale difference sequences with respect to an increasing sequence of $\sigma$-fields $\mathcal{G}_T$ generated by the history states, and satisfy the following conditions:
	\begin{align*}
		&\lim\inf_{N\rightarrow\infty}\frac{1}{T}\sum_{i=1}^{T}v_c(i)v_c(i)^\top=\mathcal{C}_1,\\
		&\lim\inf_{N\rightarrow\infty}\frac{1}{T}\sum_{i=1}^{T}v_a(i)v_a(i)^\top=\mathcal{C}_2,\\				&\lim\inf_{N\rightarrow\infty}\frac{1}{T}\sum_{i=1}^{T}v_m(i)v_m(i)^\top=\mathcal{C}_3,\\
	\end{align*}
	where $\mathcal{C}_1\in\mathbb{S}_{\succ0}^{n_{x_c}}$, $\mathcal{C}_2\in\mathbb{S}_{\succ0}^{n_{x_a}}$, and $\mathcal{C}_3\in\mathbb{S}_{\succ0}^{n_y}$ are also bounded.
\end{standing}
\begin{remark}
	Standing Assumption \ref{assump1} denotes a common requirement for analyzing the convergence of system identification algorithms, enabling the noise process to exhibit non-stationary and heavy-tailed characteristics -- see, e.g., \cite{Lai1982,PE2018,Chen1986}.
\end{remark}

The NCS-ASM in \eqref{eq:1} thus consists of two state equations and one output equation. Specifically, the first state equation represents the dynamics of the causal state variables, while the second one the dynamics of the non-causal state variables. The system output is determined by both the causal and non-causal states. Furthermore, both the causal and non-causal parts of the system are composed of multiple subsystems, and their corresponding switching sequences are different. Given some $T\in\mathbb{Z}$, which will denote the sample size of the available dataset, let the switching sequences of the causal and anti-causal parts being denoted by $\bm{s}_c\triangleq\{s_c(i)\}_{i=1}^T$ and $\bm{s}_a\triangleq\{s_a(i)\}_{i=1}^T$, respectively. Each of them corresponds to a set of parameters, i.e., $s_c(t)=i$ determines the model parameter $\theta^c_i\triangleq\{A_c(i),C_c(i),\Sigma_c(i)\}$ that is active at the time instant $t$. In particular, the sequences $\bm{s}_c$ and $\bm{s}_a$ undergo random switches with certain (fixed) probabilities over time. Let us denote the underlying switching probabilities with $\mathbb P[\bm{s}_c=i]=\pi^c_i$ and $\mathbb P[\bm{s}_a=i]=\pi^a_i$. We then have that $\sum_{i=1}^{m_c}\pi^c_i=1$ and $\sum_{i=1}^{m_a}\pi^a_i=1$. 

The complete set of model parameters that comprehensively describe the NCS-ASM can be conveniently encapsulated into a parameter object $\theta$, defined as follows: 
\[
	\theta\triangleq\left\{ \{\theta^c_i\}_{i=1}^{m_c},\{\theta^a_i\}_{i=1}^{m_a},\{\pi^c_i\}_{i=1}^{m_c},\{\pi^a_i\}_{i=1}^{m_a},\Sigma_m\right\}.
\]

\subsection{Problem statement}
Our goal is hence to estimate the \textit{unknown} model parameters $\theta$ characterizing the NCS-ASM \eqref{eq:1} with the known state dimension, number of causal system modes $m_c$ and anti-causal system modes $m_a$, together with a collection of noisy output measurements $\bm{y}$:
\[
	\bm{y}\triangleq\bm{y}_{1:T}=\{y(1),\ldots,y(T)\}.
\]

It is worth remarking that the NCS-ASM consists of both causal and non-causal parts, and their switching sequences are different. Therefore, the problem we wish to addressing has two main challenges. First, it is difficult to obtain the parameters for each subsystem of the NCS-ASM, since the states of both parts are unknown, and the system output is determined by both causal and non-causal states. Identifying both parts simultaneously is hence challenging, especially due to the continuous switching pattern of different subsystems. Second, the system has two switching sequences, and the combination of subsystems varies at different time instants. For example, at time $t$ the causal part may activate subsystem $i$ ($s_c(t)=i$) while the non-causal part may activate the subsystem $j$ ($s_a(t)=j$), thereby originating a total of $m_a\times m_c$ possible combinations. In addition, the switching  behavior of the subsystems is random and independent across different time instants:
\begin{align*}
	&\mathbb P[s_c(t)\vert s_c(t-1),\ldots,s_c(1)]=\mathbb P[s_c(t)],\\
	&\mathbb P[s_a(t)\vert s_a(t+1),\ldots,s_a(T)]=\mathbb P[s_a(t)], \quad t=1,\ldots,T.
\end{align*}

To deal with the identification problem of the NCS-ASM \eqref{eq:1}, the EM framework is adopted, which is an iterative method that can yield an estimate of the parameters at each iteration \cite{Dempster1977}. Let us denote the parameter estimate at the $k$-th iteration of the underlying algorithm as $\theta^k$. Then, the proposed method can be (qualitatively, for the moment) described by means of the following two steps:
\begin{enumerate}
	\item In the E-step, we develop a modified KF to estimate the states of the causal and anti-causal parts. Furthermore, the Bayesian rule is used to obtain a posterior estimate of the switching sequence. Subsequently, the full-data likelihood function $Q(\theta,\theta^k)$ can be calculated.
	\item In the M-step, the likelihood function $Q(\theta,\theta^k)$ is maximized with respect to the parameters $\theta$. Then, the identification of the NCS-ASM is updated, yielding $\theta^{k+1}$.
\end{enumerate}

Next section will discuss in detail each step of the proposed technique for NCS-ASM identification.

\section{The EM method for identifying NCS-ASM}\label{sec4}
By making use of the dataset $\bm{y}$, we aim at estimating the system parameters $\theta$. To this end, a standard approach is to let coincide $\hat{\theta}$, i.e., our estimate of the true $\theta$, with a maximizer of the likelihood function, namely:
\begin{equation}\label{eq:mdf}
\hat{\theta}=\underset{\theta}{\arg\max}~\ln \mathbb P_\theta(\bm{y}).
\end{equation}
where we indicate with $\mathbb P_\theta(\bm{y})$ the probability density function of the output $\bm{y}$ given some sets of parameters $\theta$. In the remainder, we will tacitly assume that the solution to the likelihood maximization problem \eqref{eq:mdf} is unique.

Given any collection of data $\bm{y}$, note that the likelihood function $\ln \mathbb P_\theta(\bm{y})$, also called marginal density function of $\bm{y}$, can be decomposed into the following form:

\begin{align*}
	&\ln \mathbb P_\theta(\bm{y})\\
	&=\ln \mathbb P_\theta[y(1)]+\sum_{i=2}^{T-1}\ln \mathbb P_\theta[y(i)]+\ln \mathbb P_\theta[y(T)]\\
	&=\ln\int\int\sum_{s_c(1)}\sum_{s_a(1)} \mathbb P_\theta[y(1)\vert x_c(1),x_a(1),s_c(1),s_a(1)]\\
	&\mathbb P_\theta[x_c(1),x_a(1),s_c(1),s_a(1)]~dx_c(1)dx_a(1)\\
	&+\ln\int\int\sum_{s_c(T)}\sum_{s_a(T)} \mathbb P_\theta[y(T)\vert x_c(T),x_a(T),s_c(T),s_a(T)]\\
	&\mathbb P_\theta[x_c(T),x_a(T),s_c(T),s_a(T)]~dx_c(T)dx_a(T)\\
	&+\sum_{i=2}^{T-1}\ln\int\int\sum_{s_c(i)}\sum_{s_a(i)} \mathbb P_\theta[y(i)\vert x_c(i),x_a(i),s_c(i),s_a(i)]\\
	&\mathbb P_\theta[x_c(i),s_c(i)\vert \bm{y}_{1:i-1})p_\theta(x_a(i),s_a(i)\vert \bm{y}_{i+1:T}]~dx_c(i)dx_a(i).
\end{align*}
Let us denote the collection of state variables over $T$ as $\bm{x}_c\triangleq\{x_c(i)\}_{i=1}^T$ and $\bm{x}_a\triangleq\{x_a(i)\}_{i=1}^T$. Recall that in the NCS-ASM \eqref{eq:1} the state variables $\bm{x}_c$ and $\bm{x}_a$ are determined by the switching sequences $\bm{s}_c$ and $\bm{s}_a$. Besides the potential nonconvexity of $\ln \mathbb P_\theta(\bm{y})$, which makes the direct maximization of $\ln \mathbb P_\theta(\bm{y})$ challenging (together with its high-dimensionality), from the decomposition above it is also clear that for calculating $\mathbb P_\theta(\bm{y})$ we need to sum over all possible values of $\bm{s}_a$, $\bm{s}_c$, thereby further complicating the solution of \eqref{eq:mdf}. 

Another way to marginalize the latent variables (such as $\bm{x}_c, \bm{x}_a, \bm{s}_c, \bm{s}_a$) is by taking the expectation over these latter. Instead of maximizing the incomplete likelihood function $\ln \mathbb P_\theta(\bm{y})$, we can estimate the conditional density of the hidden variables given the observations $\bm{y}$ and an estimate of parameter $\hat{\theta}$. Then, parameter estimate $\hat{\theta}$ can be obtained by maximizing the complete likelihood function.

To stand out our technical contributions and contrast them with existing results, we will give a sample complexity analysis related to our EM-based technique for the identification of NCS-ASM. To achieve this, we have to further postulate the following:

\begin{standing}
	The following conditions hold true:
	\begin{enumerate}
	\item The switching sequences $\bm{s}_c$, $\bm{s}_a$, and the subsystem parameters $\theta^c$, $\theta^a$ are all independent among them, i.e.,
	\begin{align*}
		\mathbb P[\bm{s}_c\vert\theta^c]&=\mathbb P[\bm{s}_c],\quad \mathbb P[\theta^c\vert \bm{s}_c]=\mathbb P[\theta^c],\\
		\mathbb P[\bm{s}_a\vert\theta^a]&=\mathbb P[\bm{s}_a],\quad \mathbb P[\theta^a\vert \bm{s}_a]=\mathbb P[\theta^a].
	\end{align*}
	\item The switching sequence follows a polynomial distribution, i.e.,
	\begin{align*}
		\mathbb P[\bm{s}_c=i]=\pi^c_i, \quad i=1,\ldots,m_c,\\
		\mathbb P[\bm{s}_a=i]=\pi^a_i, \quad i=1,\ldots,m_a,
	\end{align*}
	with $\sum_{i=1}^{m_c}\pi^c_i=1$, $\sum_{i=1}^{m_a}\pi^a_i=1$.
	\end{enumerate}
\end{standing}

Then, the full-data complete likelihood function can be expressed as follows:
\begin{align}\label{eq:1.3}
	\ln \mathbb P_{\theta}[\bm{y},\bm{x}_c,\bm{s}_c,\bm{x}_a,\bm{s}_a]= &\ln \mathbb P_\theta[\bm{y}]\nonumber\\
	&+\ln \mathbb P_\theta[\bm{x}_c,\bm{s}_c,\bm{x}_a,\bm{s}_a\vert\bm{y}].	
\end{align}
This relation directly links $\mathbb P_\theta(\bm{y})$ and $\mathbb P_\theta[\bm{y},\bm{x}_c,\bm{s}_c,\bm{x}_a,\bm{s}_a]$, with the latter depending on the unknown states $\bm{x}_c$, $\bm{x}_a$ and switching sequences $\bm{s}_c$, $\bm{s}_a$. The key step is then to approximate $\ln \mathbb P_\theta[\bm{y}]$ by the above relation \eqref{eq:1.3}, where $\bm{x}_c$, $\bm{s}_c$, $\bm{x}_a$, and $\bm{s}_a$ can be approximated by their conditional expectations based on the observed data $\bm{y}$. 
Therefore, at each iteration $k$ of our EM-based algorithm, given the estimate $\theta^k$ the conditional expectation of $\ln \mathbb P_{\theta^k}[\bm{y},\bm{x}_c,\bm{s}_c,\bm{x}_a,\bm{s}_a]$ can be obtained based on the available data $\bm{y}$ as follows:
\begin{align*}
	&\mathbb{E}_{\theta^k}[\ln \mathbb P_\theta[\bm{y},\bm{x}_c,\bm{s}_c,\bm{x}_a,\bm{s}_a]]\\
	&\qquad\qquad=\mathbb{E}_{\theta^k}[\ln \mathbb P_\theta(\bm{y})]+\mathbb{E}_{\theta^k}[\ln \mathbb P_\theta[\bm{x}_c,\bm{s}_c,\bm{x}_a,\bm{s}_a\vert\bm{y}]]\\
	&\qquad\qquad=\ln \mathbb P_\theta(\bm{y})+\mathbb{E}_{\theta^k}[\ln \mathbb P_\theta[\bm{x}_c,\bm{s}_c,\bm{x}_a,\bm{s}_a\vert\bm{y}]].
\end{align*}
Let us then define:
\begin{align*}
	Q(\theta,\theta^k)&=\mathbb{E}_{\theta^k}[{\ln \mathbb P_\theta[\bm{y},\bm{x}_c,\bm{s}_c,\bm{x}_a,\bm{s}_a]}],\\
	V(\theta,\theta^k)&=\mathbb{E}_{\theta^k}[\ln \mathbb P_\theta[\bm{x}_c,\bm{s}_c,\bm{x}_a,\bm{s}_a\vert\bm{y}]].
\end{align*}
The EM approach iteratively estimates the parameters in the following two steps. First, we compute the expectation $Q(\theta,\theta^k)$ based on $\theta^k$ obtained from the previous iteration. By the Bayesian rule and Markov properties, it can be inferred that:
\begin{align*}
	&\ln \mathbb P_\theta[\bm{y},\bm{x}_c,\bm{s}_c,\bm{x}_a,\bm{s}_a]\\
	&=\sum_{i=1}^{T}\ln \mathbb P_\theta[y(i)\vert x_c(i),x_a(i),s_c(i),s_a(i)]\\
	&+\ln \mathbb P_\theta[x_c(1),s_c(1)]+\sum_{i=2}^{T} \ln \mathbb P_\theta[x_c(i)\vert x_c(i-1),s_c(i)]\\
	&+\ln \mathbb P_\theta[x_a(T),s_a(T)]+\sum_{i=1}^{T-1} \ln \mathbb P_\theta[x_a(i)\vert x_a(i+1),s_a(i)]\\
	&=\sum_{i=1}^{T}\sum_{j=1}^{m_c}\sum_{l=1}^{m_a}\ln \mathbb P_\theta[y(i)\vert x_c(i),x_a(i)]\pi_j^c\pi_l^a\\
	&+\sum_{j=1}^{m_c}\ln \mathbb P_\theta[x_c(1)]\pi_j^c+\sum_{i=2}^{T}\sum_{j=1}^{m_c}\ln \mathbb P_\theta[x_c(i)\vert x_c(i-1)]\pi_j^c\\
	&+\sum_{l=1}^{m_a}\ln \mathbb P_\theta[x_a(T)]\pi_l^a+\sum_{i=1}^{T-1}\sum_{l=1}^{m_a}\ln \mathbb P_\theta[x_a(i)\vert x_a(i+1)]\pi_l^a.
\end{align*}
In view of the white noise assumption characterizing the disturbance affecting both state variables and measured output, note that the distribution of the these variables, given the subsystem modes $s_c(i)=j$, $s_a(i)=l$, is Gaussian too and given as follows:
\begin{align*}
	\mathbb P_\theta[y(i)\vert &x_c(i),x_a(i)]=\vert 2\pi\Sigma_m\vert^{-1/2}\\
	&\exp\{(y(i)-\mu_1(i))^\top\Sigma_m^{-1}(y(i)-\mu_1(i))\},\\
	\mathbb P_\theta[x_c(i)\vert &x_c(i-1)]=\vert 2\pi\Sigma_c(i)\vert^{-1/2}\\
	&\exp\{(x_c(i)-\mu_2(i))^\top\Sigma_c^{-1}(i)(x_c(i)-\mu_2(i))\},\\
	\mathbb P_\theta[x_a(i)\vert &x_a(i+1)]=\vert 2\pi\Sigma_a(i)\vert^{-1/2}\\
	&\exp\{(x_a(i)-\mu_3(i))^\top\Sigma_a^{-1}(i)(x_a(i)-\mu_3(i))\},
\end{align*} 
where
\begin{align*}
	\mu_1(i)&=y(i)-C_c(j)x_c(i)-C_a(l)x_a(i),\\
	\mu_2(i)&=x_c(i)-A_c(j)x_c(i-1),\\
	\mu_3(i)&=x_a(i)-A_a(l)x_a(i+1).
\end{align*}
Let us indicate with $w_{ij}^c$ the posterior probability of the switching sequence given that $s_c(i)=j$ ($w_{il}^a$ is defined similarly). Then, the objective function $Q(\theta,\theta^k)$ assumes the following form:
\begin{align}\label{eq:1.4}
	\mathbb{E}_{\theta^k}[{\ln \mathbb P_\theta[\bm{y},\bm{x}_c,\bm{s}_c,\bm{x}_a,\bm{s}_a]}\vert \bm{y}]
	=\sum_{i=1}^5 Q_i(\theta,\theta^k),
\end{align}
\begin{figure*}[!h]
	{
		\begin{equation}\label{eq:terms}
			\begin{aligned}
				Q_1(\theta,\theta^k)&=\sum_{i=1}^{T}\int\int\int\int w^c_{ij}w^a_{il}\ln \mathbb P_\theta[y(i)\vert x_c(i),x_a(i)] \mathbb P_{\theta^k}[x_c(i)\vert \bm{y}] \mathbb P_{\theta^k}[x_a(i)\vert \bm{y}]~d(x_c(i))d(x_a(i))d(\bm{s}_c)d(\bm{s}_a),\\
				Q_2(\theta,\theta^k)&=\int\int w^c_{1j}\ln \mathbb P_\theta[x_c(1)] \mathbb P_{\theta^k}[x_c(1)\vert\bm{y}]~d(x_c(1))d(\bm{s}_c),\\
				Q_3(\theta,\theta^k)&=\sum_{i=2}^{T}\int\int\int w^c_{ij}\ln \mathbb P_\theta[x_c(i)\vert x_c(i-1)] \mathbb P_{\theta^k}[x_c(i),x_c(i-1)\vert \bm{y}]~d(x_c(i))d(x_c(i-1))d(\bm{s}_c),\\
				Q_4(\theta,\theta^k)&=\int\int w^a_{Tl}\ln \mathbb P_\theta[x_a(T)] \mathbb P_{\theta^k}[x_a(T)\vert\bm{y}]~d(x_a(T))d(\bm{s}_a),\\
				Q_5(\theta,\theta^k)&=\sum_{i=1}^{T-1}\int\int\int w^a_{il}\ln \mathbb P_\theta[x_a(i)\vert x_a(i+1)] \mathbb P_{\theta^k}[x_a(i),x_c(i+1)\vert \bm{y}]~d(x_a(i))d(x_a(i+1))d(\bm{s}_a).
			\end{aligned}
		\end{equation}
	}
	\hrulefill
\end{figure*} 
where the closed form for the terms $Q_i$ is in \eqref{eq:terms}. Note that the posterior densities $w^c_{ij}$ and $w^a_{il}$, given the parameter $\theta^k$ and dataset $\bm{y}$, can be computed as follows:
\begin{align}\label{eq:posterior_c}
	w^c_{ij}= \mathbb P_{\theta^k}[s_c(i)&=j\vert\bm{y}]=\frac{\mathbb P_{\theta_i^k}[\bm{y},s_c(i)=j]}{\mathbb P_{\theta_i^k}[\bm{y}]} \nonumber\\
	&=\frac{\mathbb P_{\theta_i^k}[\bm{y}\vert s_c(i)=j]\pi^c_j}{\sum_{i=1}^{T} \mathbb P_{\theta_i^k}[\bm{y}\vert s_c(i)=j]\pi^c_j},
\end{align}
\begin{align}\label{eq:posterior_a}
	w^a_{il}= \mathbb P_{\theta^k}[s_a(i)&=l\vert\bm{y}]=\frac{\mathbb P_{\theta_i^k}[\bm{y},s_a(i)=l]}{\mathbb P_{\theta_i^k}[\bm{y}]}\nonumber\\
	&=\frac{\mathbb P_{\theta_i^k}[\bm{y}\vert s_a(i)=l]\pi^a_l}{\sum_{i=1}^{T} \mathbb P_{\theta_i^k}[\bm{y}\vert s_a(i)=l]\pi^c_l}.
\end{align}

Then, the second step is to maximize the $Q(\theta,\theta^k)$ to obtain $\theta^{k+1}$, formally defined as $\theta^{k+1}=\arg\max_\theta Q(\theta,\theta^k)$.

Algorithm~\ref{alg:EM} summarizes the two main steps of the proposed identification methodology for NCS-ASM. We characterize next the monotonic properties of the likelihood function in \eqref{eq:mdf} when the EM algorithm is iteratively applied to estimate the system parameters $\theta$:
\begin{lemma}\label{lemma:1}
Given a dataset $\bm{y}$, let $\{\theta^k\}_{k\in\mathbb{Z}}$ be the sequence generated by Algorithm~\ref{alg:EM}. Then, the likelihood function in \eqref{eq:mdf}, evaluated along $\{\theta^k\}_{k\in\mathbb{Z}}$, is non-decreasing, thereby yielding $\ln \mathbb P_{\theta^{k+1}}[\bm{y}]\geq\ln \mathbb P_{\theta^k}[\bm{y}]$ for all $k\in\mathbb{Z}$.
\end{lemma}

\begin{algorithm}[h!]
	\caption{EM-based identification of NCS-ASM}\label{alg:EM}
	\smallskip
	
	\textbf{Initialization:} Collect data $\bm y_{1:T}$, set $\theta^0$
	
	\smallskip
	
	\textbf{Iteration} $k\in\mathbb{Z}$\textbf{:}
	\smallskip
	\begin{enumerate}
		\item \textbf{E-step}: Compute $Q(\theta,\theta^k)$ using \eqref{eq:1.4}, \eqref{eq:terms}, \eqref{eq:posterior_c}, \eqref{eq:posterior_a}
		\smallskip
		\item \textbf{M-step}: Set $\theta^{k+1}=\underset{\theta}{\arg\max}~Q(\theta,\theta^k)$
	\end{enumerate}
\end{algorithm}

\section{Implementation details of the EM algorithm}\label{sec5}
We now delve into the details of the steps outlined in Algorithm~\ref{alg:EM}, ultimately establishing our main technical result characterizing the sample complexity of the proposed identification technique for NCS-ASM.

\subsection{The E-step}\label{subsec:E-step}
This step requires the calculation of the objective function $Q(\theta,\theta^k)$. Specifically, this shall be achieved on the basis of the parameter $\theta^k$ estimated in the previous iteration. Then, according to the expression of $Q(\theta,\theta^k)$ in \eqref{eq:1.4}, the expectations of states $\bm{x}_c,\bm{x}_a$ and the switching sequences $\bm{s}_c,\bm{s}_a$ given the data $\bm{y}$ are required. 

First, we calculate the posterior estimates of the switching sequences $\bm{s}_c$ and $\bm{s}_a$ by leveraging the Bayesian rule, namely $\mathbb P_\theta[\bm{s}_c\vert\bm{y}]=\mathbb P_\theta[\bm{s}_c,\bm{y}]/\mathbb P_\theta(\bm{y})$ and $\mathbb P_\theta[\bm{s}_a\vert\bm{y}]=\mathbb P_\theta[\bm{s}_a,\bm{y}]/\mathbb P_\theta(\bm{y})$. In addition, according to the formula of total probability one obtains:
\begin{align*}
	\mathbb P_\theta(\bm{y})&=\sum_{j=1}^{m_c} \mathbb P_\theta[\bm{y}\vert \bm{s}_c=j] \mathbb P[\bm{s}_c=j]=\sum_{j=1}^{m_c} \mathbb P_\theta[\bm{y}\vert \bm{s}_c=j]\pi_j^c,\\
	\mathbb P_\theta(\bm{y})&=\sum_{l=1}^{m_a} \mathbb P_\theta[\bm{y}\vert \bm{s}_a=l] \mathbb P[\bm{s}_a=l]=\sum_{l=1}^{m_a} \mathbb P_\theta[\bm{y}\vert \bm{s}_a=l]\pi_l^a.
\end{align*}
 Then, the data point can be assigned to each subsystem at time $i$ by solving the following optimization problem:
\begin{align*}
	\hat{s}_c(i)&=\underset{j\in\{1,\ldots,m_c\}}{\arg\max}~\mathbb P_\theta[y(i)\vert s_c(i)=j]\pi_j^c,\\
	\hat{s}_a(i)&=\underset{l\in\{1,\ldots,m_a\}}{\arg\max}~\mathbb P_\theta[y(i)\vert s_a(i)=l]\pi_l^a,
\end{align*}
where maximizing $\mathbb P_\theta[y(i)\vert s_c(i)=j]\pi_j^c$ is equivalent to maximizing the posterior probability of $\mathbb P_\theta[s_c(i)=j\vert y(i)]$ which is commonly used for data classification.
After obtaining $\hat{s}_c(i)$ and $\hat{s}_a(i)$, for any $(i,j)\in\{1,\ldots,T\}\times\{1,\ldots,m_c\}$ (or $(i,l)\in\{1,\ldots,T\}\times\{1,\ldots,m_a\}$), we obtained that 
\[
	w_{ij}^c= \left\{
	\begin{array}{ll}
		1 & \text{if $\hat{s}_c(i)=j$} \\
		0 & \text{else}
	\end{array}
	\right.,\quad
	w_{il}^a= \left\{
	\begin{array}{ll}
		1 & \text{if $\hat{s}_a(i)=l$} \\
		0 & \text{else}
	\end{array}
	\right.
\]

Successively, we focus on the reconstruction of the state variables $\bm{x}_c$ and $\bm{x}_a$, a task that is traditionally accomplished by means of a Kalman filter. The latter amounts to a recursive algorithm that estimates the system states by incorporating the information inferred from previous estimates and current measurements. 
Adapting the KF to our problem, however, requires few key modifications due to the following reason.
According to the system description in \eqref{eq:1}, we note that the calculation of the posterior probability $\mathbb P_{\theta}[x_c(t)\vert \bm{y}]$ is related to two equations. The first one requires to propagate the estimate from the previous time step, $\mathbb P_{\theta}[x_c(t-1)\vert \bm{y}]$, to the current time step recursively, whereas the second one involves correcting the prior estimate with the measurement equation in a reverse manner (for the non-causal states happens exactly the opposite). Then, when correcting the prior prediction of the state variables $\bm{x}_c$ and $\bm{x}_a$ using the data $\bm{y}$, we note that $\bm{x}_c$ and $\bm{x}_a$ are \textit{mutually correlated}, thereby calling for a careful design of the KF as described below. To simplify notation we omit the dependency on the switching sequence, e.g., $A_c=A_c(\hat{s}_c(t))$):

First, we need to compute the prior state estimates of $\bm{x}_c$ and $\bm{x}_a$, denoted as $\hat{\bm{x}}_c^-$ and  $\hat{\bm{x}}_a^-$, through the first two relations in \eqref{eq:1}, yielding $\hat{x}_c^-(t)=A_c\hat{x}_c(t-1)$ and $\hat{x}_a^-(t)=A_a\hat{x}_a(t+1)$. With this regard, note that the switching sequence for each step has already been calculated. Successively, the measurement equation in \eqref{eq:1} allows us to perform posterior corrections $\hat{\bm{x}}_c$ and $\hat{\bm{x}}_a$ on the underlying prior estimates $\hat{\bm{x}}_c^-$ and  $\hat{\bm{x}}_a^-$ as follows:
\begin{align*}
	\hat{x}_c(t)&=\hat{x}_c^-(t)+K_cC_c(C_c^{-1}(y(t)-C_a\hat{x}_a^-(t))-C_c\hat{x}_c^-(t))\\
	&=\hat{x}_c^-(t)+K_c(y(t)-C_a\hat{x}_a^-(t)-C_c\hat{x}_c^-(t)),\\
	\hat{x}_a(t)&=\hat{x}_a^-(t)+K_aC_a(C_a^{-1}(y(t)-C_a\hat{x}_a^-(t))-C_c\hat{x}_c^-(t))\\
	&=\hat{x}_a^-(t)+K_a(y(t)-C_a\hat{x}_a^-(t)-C_c\hat{x}_c^-(t)),
\end{align*}
where $K_c\in\mathbb{R}^{n_{x_c}\times n_y}$ and $K_a\in\mathbb{R}^{n_{x_a}\times n_y}$ are the Kalman gains for the causal and anti-causal states, respectively, whose design is critical for the effectiveness of the KF.
Before delving into the derivation of $K_c$ and $K_a$, let us first calculate the error covariance matrix for the prior state estimates based on the prior estimation errors $e_c^-(t)=x_c(t)-\hat{x}^-_c(t)$ and $e_a^-(t)=x_a(t)-\hat{x}^-_a(t)$, and the posterior estimation errors $e_c(t)=x_c(t)-\hat{x}_c(t)$ and $e_a(t)=x_a(t)-\hat{x}_a(t)$. Then, the prior error covariance and posterior estimation error covariance matrices can be defined as follows:
\begin{align*}
	P_c^-&\triangleq\mathbb{E}[(x_c(t)-\hat{x}_c^-(t))(x_c(t)-\hat{x}_c^-(t))^\top],\\
	P_a^-&\triangleq\mathbb{E}[(x_a(t)-\hat{x}_a^-(t))(x_a(t)-\hat{x}_a^-(t))^\top],\\
	P_c&\triangleq\mathbb{E}[(x_c(t)-\hat{x}_c(t))(x_c(t)-\hat{x}_c(t))^\top],\\
	P_a&\triangleq\mathbb{E}[(x_a(t)-\hat{x}_a(t))(x_a(t)-\hat{x}_a(t))^\top].\\
\end{align*}
Then, we have:
\begin{align*}
	P_c^-&=\mathbb{E}[(x_c(t)-\hat{x}_c^-(t))(x_c(t)-\hat{x}_c^-(t))^\top]\\
	&=\mathbb{E}[(A_c(x_c(t-1)-\hat{x}_c(t-1))+v_c(t))]\\
	&\hspace{2cm}(A_c(x_c(t-1)-\hat{x}_c(t-1))+v_c(t))^\top]\\
	&=A_cP_c(t-1)A_c^\top+\Sigma_c,\\	P_a^-&=\mathbb{E}[(x_a(t)-\hat{x}_a^-(t))(x_a(t)-\hat{x}_a^-(t))^\top]\\
	&=\mathbb{E}[(A_a(x_a(t+1)-\hat{x}_a(t+1))+v_a(t))\\
	&\hspace{2cm}(A_a(x_a(t+1)-\hat{x}_a(t+1))+v_a(t))^\top]\\
	&=A_aP_a(t+1)A_a^\top+\Sigma_a.
\end{align*}
Then, the Kalman gains can be calculated so that the error covariance matrices for the posterior state estimates is minimized. The posterior estimation error can be rewritten as:
\begin{align*}
e_c(t)&=x_c(t)-\hat{x}_c(t)\\
&=x_c(t)-\hat{x}_c^-(t)+K_c(y(t)-C_a\hat{x}_a^-(t)-C_c\hat{x}_c^-(t))\\
&=(I-K_cC_c)e_c^-(t)-K_cC_ae_a^-(t)-K_cv_m(t),\\
e_a(t)&=x_a(t)-\hat{x}_a(t)\\
&=x_a(t)-\hat{x}_a^-(t)+K_a(y(t)-C_a\hat{x}_a^-(t)-C_c\hat{x}_c^-(t))\\
&=(I-K_aC_a)e_a^-(t)-K_aC_ce_c^-(t)-K_av_m(t),
\end{align*}
while the error covariance matrices of the state estimates:
\begin{align}
	P_c&=\mathbb{E}[((I-K_cC_c)e_c^-(t)-K_cC_ae_a^-(t)-K_cv_m(t))\nonumber\\
	&\hspace{1.2cm}((I-K_cC_c)e_c^-(t)-K_cC_ae_a^-(t)-K_cv_m(t))^\top]\nonumber\\
	&=(I-K_cC_c)P_c^-(I-K_cC_c)^\top+K_cC_aP_a^-C_a^\top K_c^\top\nonumber\\
	&\hspace{6.2cm}+K_c\Sigma_mK_c^\top\nonumber\\
	&=P_c^--P_c^-C_c^\top K_c^\top-K_cC_cP_c^-+K_cC_cP_c^-C_c^\top K_c^\top\nonumber\\
	&\hspace{3.3cm}+K_cC_aP_a^-C_a^\top K_c^\top+K_c\Sigma_mK_c^\top, \label{eq:1.5.1}\\
	P_a&=\mathbb{E}[((I-K_aC_a)e_a^-(t)-K_aC_ce_c^-(t)-K_av_m(t))\nonumber\\
	&\hspace{1.2cm}((I-K_aC_a)e_a^-(t)-K_aC_ce_c^-(t)-K_av_m(t))^\top]\nonumber\\
	&=(I-K_aC_a)P_a^-(I-K_aC_a)^\top+K_aC_cP_c^-C_c^\top K_a^\top\nonumber\\
	&\hspace{6.2cm}+K_a\Sigma_mK_a^\top\nonumber\\
	&=P_a^--P_a^-C_a^\top K_a^\top-K_aC_aP_a^-+K_aC_aP_a^-C_a^\top K_a^\top\nonumber\\
	&\hspace{3.3cm}+K_aC_cP_c^-C_c^\top K_a^\top+K_a\Sigma_mK_a^\top, \label{eq:1.5.2}
\end{align}
where the second equality in each derivation is established based on the independence of $e_c^-(t)$, $e_a^-(t)$ and $v_m(t)$. Note that minimizing the variances of $P_c$ and $P_a$ is equivalent to minimizing their traces. Therefore, given the unconstrained nature of such trace minimization, the optimal Kalman gains $K_c$ and $K_a$ can be found as:
\begin{align*}
	\frac{d(\textrm{tr}(P_c))}{dK_c}&=0-2[C_cP_c^-]^\top+2K_cC_cP_c^-C_c^\top\\
	&\hspace{2.2cm}+2K_cC_aP_a^-C_a^\top+2K_c\Sigma_m=0,\\
	\Rightarrow K_c&=(C_cP_c^-C_c^\top+C_aP_a^-C_a^\top+\Sigma_m)^{-1}(P_c^-C_c^\top),\\
	\frac{d(\textrm{tr}(P_a))}{dK_a}&=0-2[C_aP_a^-]^\top+2K_aC_aP_a^-C_a^\top\\
	&\hspace{2.2cm}+2K_aC_cP_c^-C_c^\top+2K_a\Sigma_m=0,\\
	\Rightarrow K_a&=(C_aP_a^-C_a^\top+C_cP_c^-C_c^\top+\Sigma_m)^{-1}(P_a^-C_a^\top).
\end{align*}

By substituting the Kalman gains above into \eqref{eq:1.5.1}--\eqref{eq:1.5.2}, the updated error covariance matrices can be obtained as:
\begin{align*}
	P_c&=P_c^--P_c^-C_c^\top K_c^\top-K_cC_cP_c^-+P_c^-C_c^\top K_c^\top\\
	&=(I-K_cC_c)P_c^-,\\
	P_a&=P_a^--P_a^-C_a^\top K_a^\top-K_aC_aP_a^-+P_a^-C_a^\top K_a^\top\\
	&=(I-K_aC_a)P_a^-.
\end{align*}
By completing the steps of the modified KF, including the prediction, measurement update, and error covariance matrix update \cite{Angelo2010,Kalman}, one can obtain all the posterior estimates of the state vectors $\bm{x}_c$ and $\bm{x}_a$, which are optimal state estimates based on the available measurements and prior knowledge. In addition, to ensure the convergence of the proposed state estimation method, we establish the following properties of the state estimates:

\begin{lemma}\label{lemma:boundendness}
	Let $\eta_c(t)=x_c(t)-A(\hat{s}_c(t))x_c(t-1)$, $\eta_a(t)=x_a(t)-A(\hat{s}_a(t))x_a(t+1)$, and $\eta_m(t)=y(t)-C_c(\hat{s}_c(t))x_c(t)-C_a(\hat{s}_a(t))x_a(t)$. There exist $\alpha_1$, $\alpha_2$, $\alpha_3 > 0$ so that $\|\eta_c(t)\|^2\le\alpha_1$, $\|\eta_a(t)\|^2\le\alpha_2$, and $\|\eta_m(t)\|^2\le\alpha_3$, for all $t\in\mathbb{Z}$.
\end{lemma}

Lemma \ref{lemma:boundendness} states that the error of state estimation is bounded in the mean square sense, regardless of how the state trajectory evolves in time.
 
By making use of the steps described in this subsection one is able to find an expression for the objective function $Q(\theta,\theta^k)$. We will discuss next how to actually maximize it with respect to its first argument.

\subsection{The M-step}
The second step in Algorithm~\ref{alg:EM} requires the maximization of $Q(\theta,\theta^k)$ to update the parameters estimate $\theta^k$:
\[
	\theta^{k+1}=\underset{\theta}{\arg\max}~Q(\theta,\theta^k).
\]
Let us first focus on the elements $\{\{\pi^c_i\}_{i=1}^{m_c},\{\pi^a_i\}_{i=1}^{m_a}\}$, and recall that the objective function reads as:
\begin{align*}
	&Q(\theta,\theta^k)\\
	&=\mathbb{E}_{\theta^k}[{\ln \mathbb P_\theta[\bm{y},\bm{x}_c,\bm{s}_c,\bm{x}_a,\bm{s}_a]}\vert \bm{y}],\\
	&=\int\int\int\int \ln \mathbb P_\theta[\bm{y},\bm{x}_c,\bm{s}_c,\bm{x}_a,\bm{s}_a] \mathbb P_{\theta^k}[\bm{x}_c\vert\bm{y}] \mathbb P_{\theta^k}[\bm{x}_a\vert\bm{y}]\\
	&\hspace{2.5cm} \mathbb P_{\theta^k}[\bm{s}_c\vert\bm{y}] \mathbb P_{\theta^k}[\bm{s}_a\vert\bm{y}]~d(\bm{x}_c)d(\bm{x}_a)d(\bm{s}_c)d(\bm{s}_a)\\
	&=\sum_{i=1}^{T}\sum_{j=1}^{m_c}\sum_{l=1}^{m_a}\int\int w^c_{ij}w^a_{il} \ln \mathbb P_\theta[y(i)\vert x_c(i),x_a(i)]\\
	&\hspace{1.7cm} \mathbb P_{\theta^k}[x_c(i)\vert \bm{y}] \mathbb P_{\theta^k}[x_a(i)\vert \bm{y}]\pi^c_j\pi^a_l~d(x_c(i))d(x_a(i)).
\end{align*}
The $(k+1)$-th estimate of $\{\{\pi^c_i\}_{i=1}^{m_c},\{\pi^a_i\}_{i=1}^{m_a}\}$ can hence be obtained in closed-form by applying the first-order optimality conditions as follows:
\begin{align*}
	\pi^c_j&=\underset{\pi^c_j}{\arg\max}~Q(\theta,\theta^k)=\frac{\sum_{i=1}^{T}w^c_{ij}}{\sum_{i=1}^{T}\sum_{j=1}^{m_c}w^c_{ij}},\\	\pi^a_l&=\underset{\pi^a_l}{\arg\max}~Q(\theta,\theta^k)=\frac{\sum_{i=1}^{T}w^a_{il}}{\sum_{i=1}^{T}\sum_{l=1}^{m_a}w^a_{il}}.
\end{align*}
Furthermore, also the expression for the parameters $\{ \{\theta^c_i\}_{i=1}^{m_c},\{\theta^a_i\}_{i=1}^{m_a},\Sigma_m\}$ can be computed in closed-form by using the switching least-squares approach as follows:
\begin{align*}
	&A_c(j)=\underset{A_c(j)}{\arg\min}~\sum_{i=1}^{T}w^c_{ij}\Vert x_c(i)-A_c(j)x_c(i-1)\Vert^2,\\
	&A_a(l)=\underset{A_a(l)}{\arg\min}~\sum_{i=1}^{T}w^a_{il}\Vert x_a(i)-A_a(l)x_a(i+1)\Vert^2,\\
	&(C_c(j),C_a(l))=\\
	&\underset{(C_c(j),C_a(l))}{\arg\min}~\sum_{i=1}^{T}w^c_{ij}w^a_{il}\Vert y(i)
	-C_c(j)x_c(i)-C_a(l)x_a(i)\Vert^2.
\end{align*}
Then, the covariance matrices related to the disturbances $v_c$, $v_a$, and $v_m$ can also be estimated as:
{
\begin{align*}
	&\Sigma_c(j)=\\
	&\sum_{i=1}^{T} w^c_{ij}(x_c(i)-A_c(j)x_c(i-1))(x_c(i)-A_c(j)x_c(i-1))^\top,\\
	&\Sigma_a(l)=\\
	&\sum_{i=1}^{T}w^a_{il}(x_a(i)-A_a(l)x_a(i+1))(x_a(i)-A_a(l)x_a(i+1))^\top,\\
	&\Sigma_m=\sum_{i=1}^{T} w^c_{ij}w^a_{il}(y(i)-C_c(j)x_c(i)-C_a(l)x_a(i))\\
	&\hspace{3.4cm}(y(i)-C_c(j)x_c(i)-C_a(l)x_a(i))^\top.
\end{align*}
}
To show the convergence rate of the system matrices, we need the following definition of strong consistency of the parameter estimates, and few auxiliary results. Recall that $\hat \theta$ is the estimate of $\theta$ made by exploiting $T$ samples. 
\begin{myDef}\textup{(\cite{Lai1982})} The estimate $\hat{\theta}$ is strongly consistent if 
	$
		\lim_{T\rightarrow\infty}\hat{\theta}=\theta.
	$
\end{myDef}

\begin{lemma}\textup{(\cite[Lemma~3]{Sayedana2024})}\label{lemma3}
	The following asymptotic relations hold true almost surely (a.s.):
	\begin{align*}
		&\left\lVert\sum_{i=1}^{T}A(s_c(i))x_c(i)v_c^\top(i)+v_c(i)x_c^\top(i)A(s_c(i))\right\rVert=o(T),\\
		&\left\lVert \sum_{i=1}^{T}A(s_a(i))x_a(i)v_a^\top(i)+v_a(i)x_a^\top(i)A(s_a(i))\right\rVert=o(T).
	\end{align*}
\end{lemma}
\begin{lemma}\textup{(\cite{Lai1982})}\label{lemma1}
	The standard least-squares solution can be expressed as
	$
		\hat{\beta}_T=\arg\min_{\beta}\Vert z-\beta^\top u\Vert^2,
	$
	where $u$ is the system inputs, $z$ is the system output, $\beta$ is the unknown parameters. Let $U_T=\sum_{i=1}^{T}u(i)u(i)^\top$. If
	\begin{enumerate}
		\item[(C1)] $\lambda_{\textrm{min}}(U_T)\rightarrow\infty$ a.s., and
		\item[(C2)] $\log\lambda_{\textrm{max}}(U_T)=o(\lambda_{\textrm{min}}(U_T))$ a.s.,
	\end{enumerate}
	then the least-squares estimate $\hat{\beta}_T$ is strongly consistent with convergence rate
	\[
		\Vert \hat{\beta}_T-\beta\Vert_\infty=\mathcal{O}\left(\sqrt{\frac{\log(\lambda_{\textrm{max}}(U_T))}{\lambda_{\textrm{min}}(U_T)}}\right)~a. s.
	\]
\end{lemma}
Lemma \ref{lemma1} indicates that the convergence rate of $\hat{\beta}_T$ depends on the covariance matrix of the system inputs.

We are now ready to establish the convergence rate for $\hat{\theta}$. Due to the possible different active subsystems at time $t$, it is convenient to define the following partition of the considered time interval $\{1, \ldots, T\}$ as $\mathbb{T}^c_{j,T}=\{t\leq T\vert s_c(t)=j\}$ and $\mathbb{T}^a_{l,T}=\{t\leq T\vert s_a(t)=l\}$.
\begin{thm}\label{th3}
	Let $W^c_{j,T}=\sum_{t\in\mathbb{T}^c_{j,T}} x_c(t)x_c^\top(t)$ and $W^a_{l,T}=\sum_{t\in\mathbb{T}^a_{l,T}}x_a(t)x_a^\top(t)$. Then, the estimate $\hat{\theta}$ generated by Algorithm~\ref{alg:EM} is strongly consistent for any $\bm{s}_c\in\Lambda_c^T$ and $\bm{s}_a\in\Lambda_a^T$, and the convergence rates are
	\begin{align*}
		&\Vert \hat{A}_c(j)-A_c(j)\Vert_\infty\leq\mathcal{O}\left(\sqrt{\frac{\log(\lambda_{\textrm{max}}(W^c_{j,T}))}{\lambda_{\textrm{min}}(W^c_{j,T})}}\right),\\
		&\Vert \hat{A}_a(l)-A_a(l)\Vert_\infty\leq\mathcal{O}\left(\sqrt{\frac{\log(\lambda_{\textrm{max}}(W^a_{l,T}))}{\lambda_{\textrm{min}}(W^a_{l,T})}}\right),\\
		&\Vert \hat{C}_c(j)-C_c(j)\Vert_\infty\leq\mathcal{O}\left(\sqrt{\frac{\log(\lambda_{\textrm{max}}(W^c_{j,T}))}{\lambda_{\textrm{min}}(W^c_{j,T})}}\right),\\
		&\Vert \hat{C}_a(l)-C_a(l)\Vert_\infty\leq\mathcal{O}\left(\sqrt{\frac{\log(\lambda_{\textrm{max}}(W^a_{l,T}))}{\lambda_{\textrm{min}}(W^a_{l,T})}}\right).\\
	\end{align*}
	Furthermore, the convergence rate of $\hat{\Sigma}_c(j), \hat{\Sigma}_a(l), \hat{\Sigma}_m$ are
	\begin{align*}
		&\Vert \hat{\Sigma}_c(j)-\Sigma_c(j)\Vert_\infty\leq\mathcal{O}\left(\frac{\log(T)}{T}\right),\\
		&\Vert \hat{\Sigma}_a(l)-\Sigma_a(l)\Vert_\infty\leq\mathcal{O}\left(\frac{\log(T)}{T}\right),\\
		&\Vert \hat{\Sigma}_m-\Sigma_m\Vert_\infty\leq\mathcal{O}\left(\frac{\log(T)}{T}\right).
	\end{align*}
\end{thm}
\begin{remark}
	Theorem \ref{th3} gives data-dependent upper bounds for the estimation errors of the parameter matrices. In order to have a data-independent characterization of the convergence rate for adaptive control or reinforcement learning purposes, in the proof of Theorem~\ref{th3}, specifically equation \eqref{eq6.1}, we provide with the corresponding convergence rate of the parameter estimate $\hat{\theta}$, which is equal to $\mathcal{O}(\sqrt{\log(T)/T})$.
\end{remark}

\section{Numerical examples}\label{sec6}
We now verify the effectiveness of the proposed methodology on two simulation examples. In both cases, we note that the true switching sequences $\bm{s}_c$ and $\bm{s}_a$ are only used to verify the accuracy of the estimated switching sequences, i.e., $\hat{\bm{s}}_c$ and $\hat{\bm{s}}_a$. As performance index we make use of the mode match rate, defined as:
\[
	L_{\text{mr}}= \frac{1}{T}\sum_{i=1}^{T}\iota(s_c(i),\hat{s}_c(i)),
\]
where $\iota(\cdot,\cdot)$ denotes the standard indicator function, i.e., $\iota(s_c(i),\hat{s}_c(i))=1$ if $s_c(i)=\hat{s}_c(i)$, $0$ otherwise. 
\subsection{Example 1: Academic NCS-ASM}
For illustrative purposes, we start by considering a simple non-causal system described in \eqref{eq:1} with $m_c=m_a=2$ modes and main parameters reported in Table~\ref{tab:1} (refer to the ``True'' columns). The dimensions of the outputs, causal states, and anti-causal states are $n_y=1,n_{x_c}=n_{x_a}=2$. The probabilities of all modes are $\pi^c_1=0.7,\pi^c_2=0.3,\pi^a_1=\pi^a_2=0.5$. The system is excited with white noise with zero mean and finite variance, and the data length is $T=10^4$.

\begin{figure}
	\centering
		\includegraphics[height=0.8\linewidth]{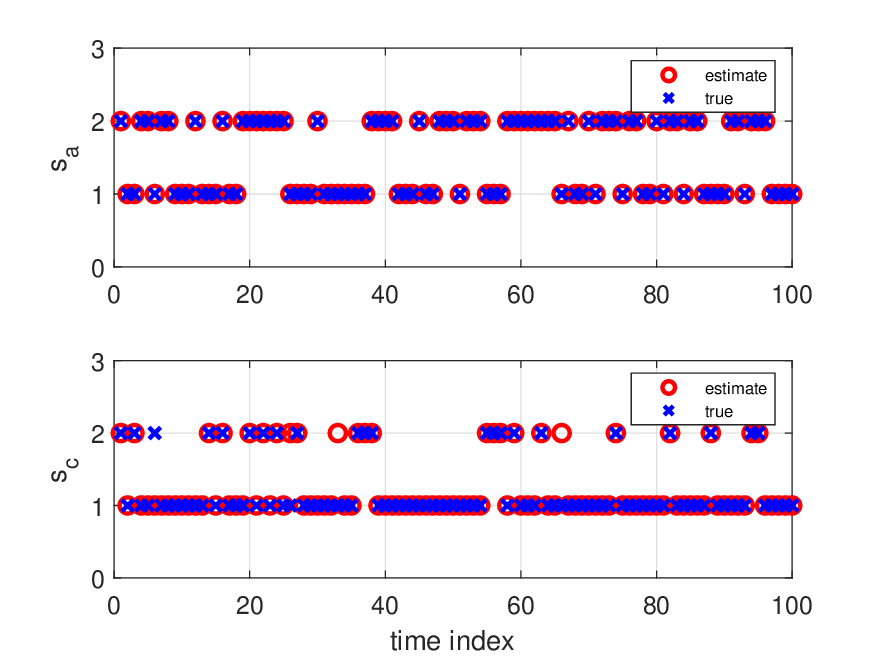} 
		\caption{The true (blue cross) and estimated (red circle) mode sequences over a certain time window of length $100$.}
		\label{fig:1}
\end{figure}

\begin{figure}
	\centering
	\includegraphics[height=0.8\linewidth]{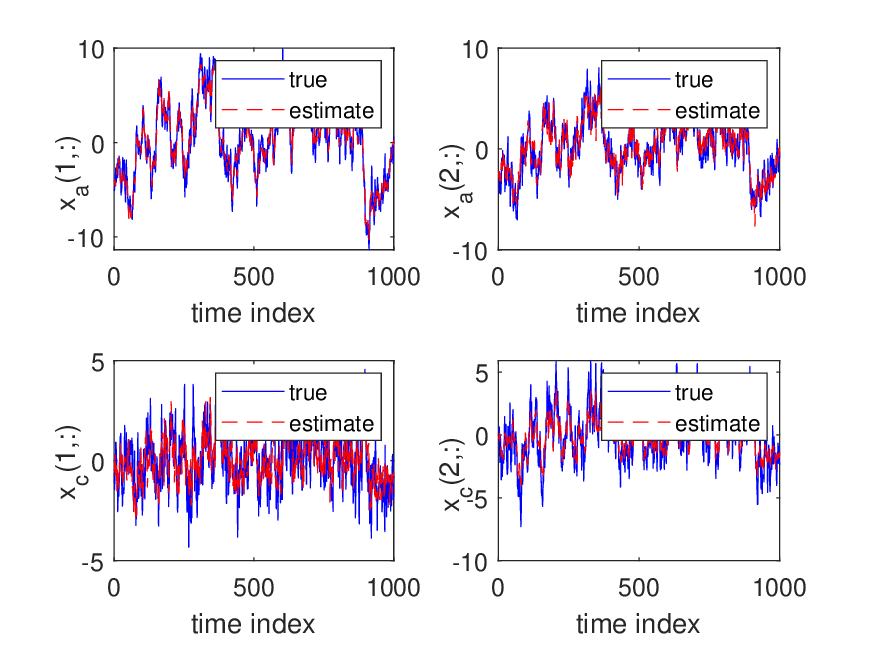} 
	\caption{Dynamical evolution of the true state variables $\bm{x}_c$ and $\bm{x}_a$ (solid blue line), and of the estimated ones $\hat{\bm x}_c$, and $\hat{\bm x}_a$ (dashed red lines).}
	\label{fig:2}
\end{figure}

The true and estimated parameters are reported in Table \ref{tab:1}, which clearly shows that the parameter estimates are very close to their true values. In Fig. \ref{fig:1} we report the partial estimation of the switching sequences $\bm{s}_c$ and $\bm{s}_a$, where the mode match rates are $97.4\%$ and $99.2\%$, respectively. Note that our method achieves an accurate parameter estimate, since each data point can be accurately assigned to the corresponding mode. 
To better validate the accuracy of the proposed algorithm in parameter estimation, Fig. \ref{fig:2} illustrates the estimated states using the modified KF. The relative estimation errors, defined as $\delta_c=\Vert x_c-\hat{x}_c\Vert^2/\Vert x_c\Vert^2$ ($\delta_a$ has the same structure), are $\delta_c=3.74\%$ and $\delta_a=3.14\%$, respectively.

\begin{table*}[!hbtp]
	\centering
	\renewcommand{\arraystretch}{1.2}
	\caption{The true and estimated system parameters}
	\label{tab:1}
	\begin{tabular}{|c|c|c|c|c|c|}\hline
		& True & Estimate &      & True & Estimate\\\hline
		$A_a(1)$&$\begin{bmatrix} 1 & 0 \\ 0 & 1 \end{bmatrix}$&    $\begin{bmatrix} 0.9681 & 0.0120 \\ 0.0142 & 0.9868 \end{bmatrix}$&$A_a(2)$&$\begin{bmatrix} 0.6 & 0.2 \\ 0.3 & 0.8 \end{bmatrix}$&$\begin{bmatrix} 0.6242 & 0.1992 \\ 0.3283 & 0.7738 \end{bmatrix}$\\\hline $A_c(1)$&$\begin{bmatrix} 1 & 0.2 \\ 0.3 & 0.8 \end{bmatrix}$&$\begin{bmatrix} 1.0131 & 0.2130 \\ 0.2849 & 0.8333 \end{bmatrix}$&$A_c(2)$&$\begin{bmatrix} 0.8 & 0.2 \\ 0.3 & 0.5 \end{bmatrix}$&$\begin{bmatrix} 0.8118 & 0.1899 \\ 0.3291 & 0.4784 \end{bmatrix}$\\\hline
		$C_a(1)$&$\begin{bmatrix} 0.2 & 0.6 \end{bmatrix}$&    $\begin{bmatrix} 0.2011 & 0.5962\end{bmatrix}$&$C_a(2)$&$\begin{bmatrix} 0.3 & 0.76 \end{bmatrix}$&$\begin{bmatrix} 0.2850 & 0.7677 \end{bmatrix}$\\\hline
		$C_c(1)$&$\begin{bmatrix} 0.3 & 0.7 \end{bmatrix}$&    $\begin{bmatrix} 0.2983 & 0.6979\end{bmatrix}$&$C_c(2)$&$\begin{bmatrix} 0.7 & 0.2 \end{bmatrix}$&$\begin{bmatrix} 0.7023 & 0.2029 \end{bmatrix}$\\\hline
		$\pi^c_1$&0.7&0.6963&$\pi^c_2$&0.3&0.3037\\\hline
		$\pi^a_1$&0.5&0.493&$\pi^a_2$&0.5&0.507\\\hline
		$\Sigma_a(1)$&$\begin{bmatrix} 1 & 0 \\ 0 & 1 \end{bmatrix}$&    $\begin{bmatrix} 1.1111 & -0.0711 \\ -0.0711 & 0.9865 \end{bmatrix}$&$\Sigma_a(2)$&$\begin{bmatrix} 1 & 0 \\ 0 & 1 \end{bmatrix}$&$\begin{bmatrix} 0.9307 & 0.0567 \\ 0.0567 & 1.0386 \end{bmatrix}$\\\hline
		$\Sigma_c(1)$&$\begin{bmatrix} 1 & 0 \\ 0 & 1 \end{bmatrix}$&    $\begin{bmatrix} 0.9773 & -0.0067 \\ -0.0067 & 0.9763 \end{bmatrix}$&$\Sigma_c(2)$&$\begin{bmatrix} 1 & 0 \\ 0 & 1 \end{bmatrix}$&$\begin{bmatrix} 1.0134 & -0.0001 \\ -0.0001 & 0.9850 \end{bmatrix}$\\\hline
		$\Sigma_m$&1&1.0049&&&\\\hline
	\end{tabular}
\end{table*} 

For comparison purposes, we now implement the EM method for jump Markov linear systems proposed in \cite{Mark2022}. The length of the data is set to $T=10^4$. The transition matrix in \cite{Mark2022} is set to $\mathcal{T}=\left[\begin{smallmatrix}
	0.5&0.5\\
	0.5&0.5
\end{smallmatrix}\right]$, and the probability of the switching sequence in this paper is set to $\pi^c_1=\pi^c_2=\pi^a_1=\pi^a_2=0.5$. The subsystem match rates of the proposed method and \cite{Mark2022} are compared at different noise levels by assuming $\Sigma=\Sigma_c=\Sigma_a$. The identification accuracy of the switching sequences are shown in Table~\ref{tab:2}.
\begin{table}[!hbtp]
	\centering
	\caption{The mode match rates achieved by the EM algorithm in \cite{Mark2022} and by the proposed method.}
	\label{tab:2}
	\begin{tabular}{ccccc}\hline
		  & $L_{\text{mr}}(\bm{s})$ \cite{Mark2022} &  $L_{\text{mr}}(\bm{s}_c)$    & $L_{\text{mr}}(\bm{s}_a)$ \\\hline
		$\Sigma=0.00$& $100\%$&$100\%$ &$100\%$ &\\\hline 
		$\Sigma=0.01$&$99.5\%$ & $98.5\%$& $99.3\%$&\\\hline 
		$\Sigma=0.1$& $96.5\%$&$97.6\%$ &$99.1\%$ &\\\hline 
		$\Sigma=1$& $89.2\%$&$97.4\%$ &$99.2\%$ &\\\hline 
	\end{tabular}
\end{table} 

\begin{figure}
	\centering
	\includegraphics[height=0.8\linewidth]{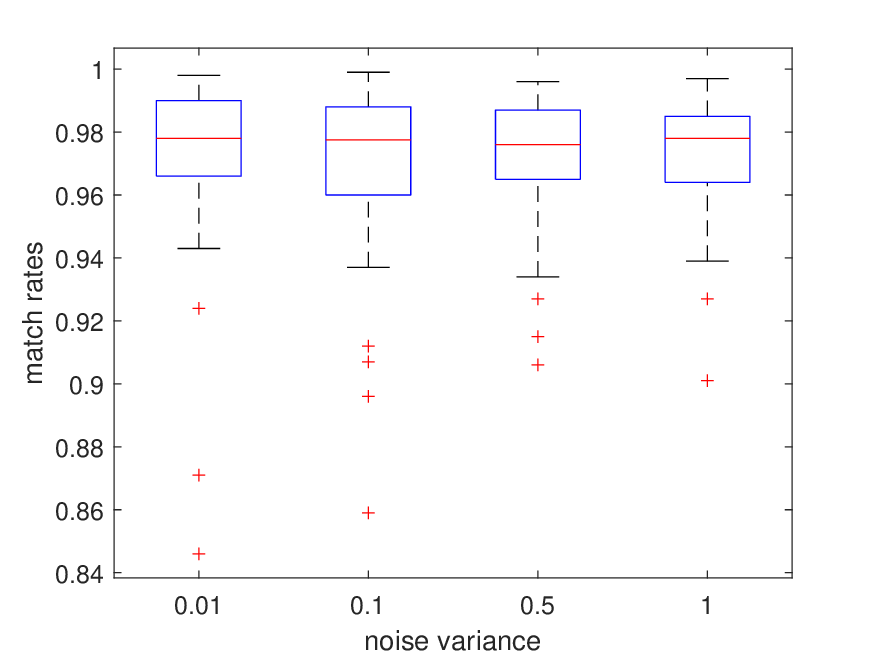} 
	\caption{Match rates obtained by the proposed algorithm for different noise levels.}
	\label{fig:3}
\end{figure}

To verify the robust performance of the proposed method against several noise levels, we run $100$ Monte Carlo experiments under four different noise conditions, i.e., $\Sigma\in\{0.01I,0.1I,0.5I,I\}$. In Fig. \ref{fig:3} we report the mean and the variance of the match rates in all the considered cases. We observe that the estimation accuracy of the switching sequence is not significantly affected by the noise variance, since even for high noise levels the estimation accuracy can still reach $98\%$ due to the excellent performance of the modified KF.

\subsection{Example 2: The Department Store Inventory Price Index }
In this subsection we adopt ``The Department Store Inventory Price Index"(DSIP) dataset from  The Bureau of Labor Statistics (BLS). These data come from inventory weighted price indices of goods carried by department stores.  

The department store inventory product prices can be influenced by various factors, such as supply and demand dynamics, market competition, seasonal variations and so on. These factors can collectively contribute to a complex relationship with the prices, thereby, suggesting that the inventory product prices may adhere to a mixed causal non-causal system. There are many causal relationships that affect the prices of department stores, such as supply and demand, cost of production, and market competition. These will directly lead to changes of future prices. There are also many non-causal factors that may not directly cause changes in the price of goods, such as economic conditions, consumer preferences, and trends. More importantly, seasonal changes can also lead to changes in its price index. Therefore, the system dynamics may change at different times to characterize DSIP. To sum up, a NCS-ASM  model \eqref{eq:1} is suitable for describing the DSIP.

\begin{figure*}
	\centering
\begin{subfigure}{0.32\linewidth}
	\includegraphics[height=0.6\linewidth]{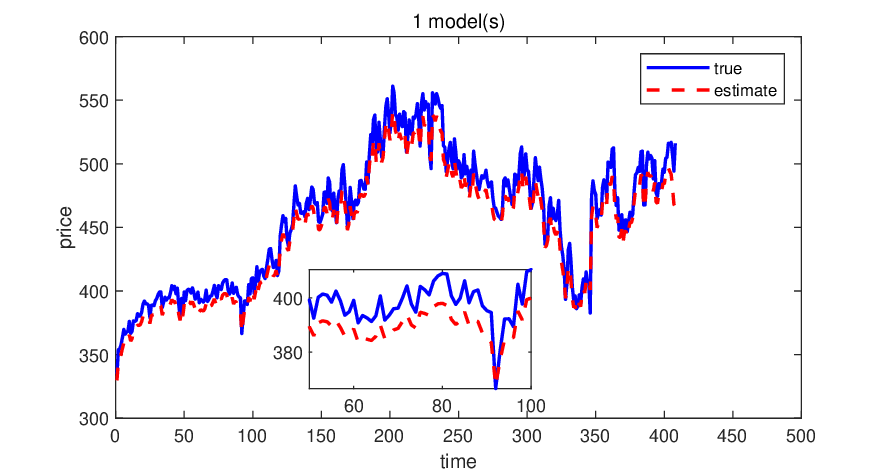} 
	\caption{}
\end{subfigure}
	\centering
\begin{subfigure}{0.32\linewidth}
		\includegraphics[height=0.6\linewidth]{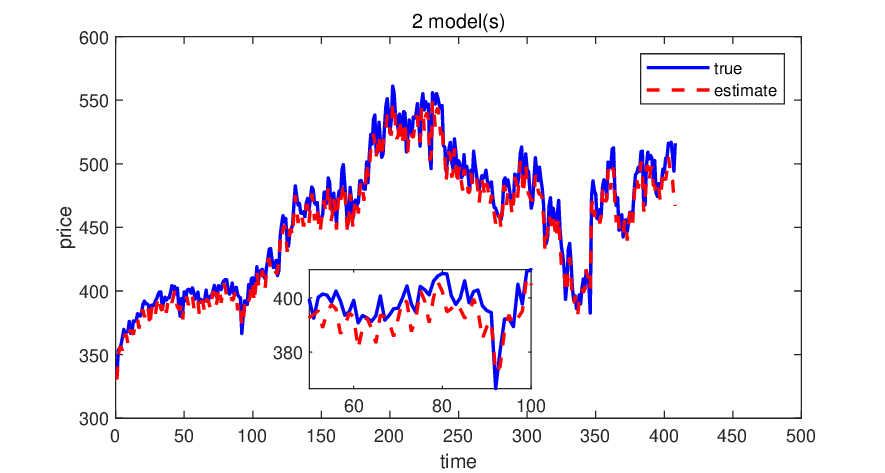} 
	\caption{}
\end{subfigure}
	\centering
\begin{subfigure}{0.32\linewidth}
	\includegraphics[height=0.6\linewidth]{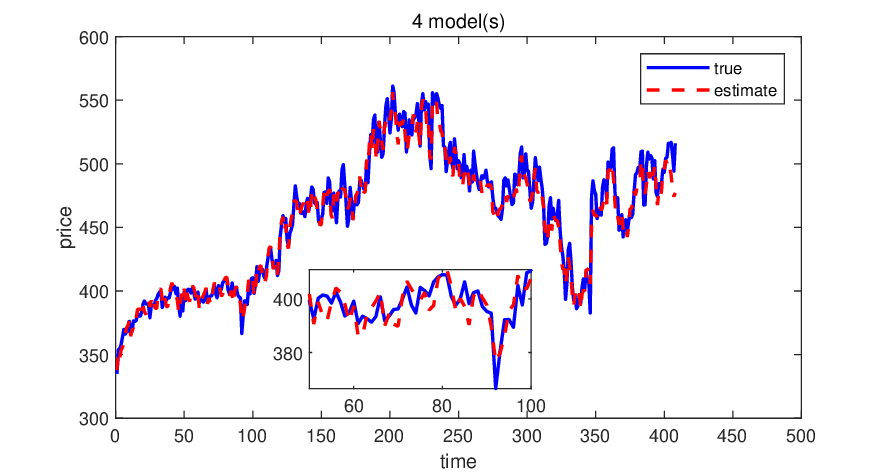} 
	\caption{}
\end{subfigure}
	\centering
	\caption{The estimated prices and the true prices with different numbers of the subsystem. (a) $\bm{s}_c=\bm{s}_a$ and $m_c=m_a=1$; (b) $\bm{s}_c\neq \bm{s}_a$ and $m_c=m_a=1$; (c) $\bm{s}_c\neq \bm{s}_a$ and $m_c=m_a=2$}\label{fig:5}
\end{figure*}

In Fig. \ref{fig:5} we show the true prices and the estimated prices with different number of subsystems. Specifically, we can infer that the larger the number of subsystems, the better the ability to describe the changes in the commodity price index. The estimation errors $\delta=\Vert \bm{y}-\hat{\bm{y}}\Vert/\Vert \bm{y}\Vert$ with different number of subsystems are shown in Table \ref{tab:3}.
\begin{table}[!hbtp]
	\centering
	\caption{The estimation errors against different number of subsystems.}
	\label{tab:3}
	\begin{tabular}{cccc}\hline
		switching sequence & $\#$ of $\bm{s}_c$ &  $\#$ of $\bm{s}_a$& $\delta$  \\\hline
		$\bm{s}_c=\bm{s}_a$& $m_c=1$&$m_a=1$&0.0249 \\\hline 
		$\bm{s}_c\neq \bm{s}_a$& $m_c=1$&$m_a=1$&0.0195 \\\hline
		$\bm{s}_c\neq \bm{s}_a$& $m_c=2$&$m_a=2$& 0.0188\\\hline
	\end{tabular}
\end{table} 

In conclusion, from Fig. \ref{fig:5} and Table \ref{tab:3} we note that switching systems with a larger number of modes can better describe the variations in DSIP, because seasonal changes can lead to different patterns in dynamic systems. The proposed identification method can therefore accurately capture the typical fluctuations of the DSIP.

\section{Conclusion}\label{sec8}
We have proposed an expectation-maximization framework for identifying non-causal systems with arbitrary switching modes. In the E-step, we have embedded the reconstructed switching sequence into the modified Kalman filter so that the proposed algorithm can handle the joint state variable estimation for the causal and anti-causal parts. Furthermore, in M-step we have developed a switching least-squares algorithm  that can get the parameter estimates in closed-form. From a technical perspective, we have established the convergence of our identification methodology, also deriving an upper bound $\mathcal{O}(\sqrt{\log(T)/T})$ for the parameter errors. 

Note that the identification algorithm proposed in this paper can be adapted to the identification of switching linear descriptor systems with minor modifications, since a descriptor state-space model can be represented in the mixed causal and anti-causal form. When the subsystems are nonlinear, however, the identification task becomes more challenging, thus posing greater difficulties. This aspect will be further investigated in our future work. In addition, addressing the joint identification of structured subsystems and piecewise constant switching sequences is an interesting future research direction.



\begin{thebibliography}{99}     

\bibitem{Tan2023} K. Tan, W. J. Parquette, \& M. Tao. (2023). A predictive algorithm for maximum power point tracking in solar photovoltaic systems through load management. Solar Energy, 265, 112127.
\bibitem{Liao2024} S. Liao, Y. Wu, K. Ma, \& Y. Niu, (2024). Ant Colony Optimization With Look-Ahead Mechanism for Dynamic Traffic Signal Control of IoV Systems. IEEE Internet of Things Journal, 11(1), 366-377.
\bibitem{Carloni2007} R. Carloni, R. G. Sanfelice, A. R. Teel, \& C. Melchiorri. (2007). A hybrid control strategy for robust contact detection and force regulation. In Proc. American Control Conf., New York City, USA, 1461-1466.
\bibitem{Schlegl2003} T. Schlegl, M. Buss, \& G. Schmidt. (2003). A hybrid systems approach toward modeling and dynamical simulation of dextrous manipulation. IEEE/ASME Trans. on Mechatronics, 8(3), 352-361.




\bibitem{Chan2008} Chan, A. B., \& Vasconcelos, N. (2008). Modeling, clustering, and segmenting video
with mixtures of dynamic textures. IEEE Transactions on Pattern Analysis and
Machine Intelligence, 30(5), 909–926.

\bibitem{Ferrari2003} Ferrari-Trecate, G., Muselli, M., Liberati, D., \& Morari, M. (2003). A clustering
technique for the identification of piecewise affine systems. Automatica, 39(2),
205–217.

\bibitem{Garulli2012} Garulli, Andrea, Paoletti, Simone, \& Vicino, Antonio (2012). A survey on switched
and piecewise affine system identification. In 16th IFAC symposium on system
identification, Brussels, Belgium (pp. 344–355).

\bibitem{Bianchi2021} Bianchi, Federico, Breschi, Valentina, Piga, Dario, \& Piroddi, Luigi (2021). Model
structure selection for switched NARX system identification: A randomized
approach. Automatica, 125, Article 109415.

\bibitem{Mark2022} Mark P. Balenzuela, Adrian G. Wills, Christopher Renton, \& Brett Ninness. (2022). Parameter estimation for Jump Markov Linear Systems. Automatica, 135 109949.

\bibitem{Bemporad2018} Alberto Bemporad, Valentina Breschi, Dario Piga, \& Stephen P. Boyd. (2018). Fitting jump models. Automatica, 96, 11-21.

\bibitem{Piga2020} Dario Piga, Valentina Breschi, \& Alberto Bemporad. (2020). Estimation of jump Box–Jenkins models. Automatica, 120 109126.

\bibitem{Sayedana2024} Borna Sayedana, Mohammad Afshari, Peter E. Caines, \& Aditya Mahajan. (2024). Strong Consistency and Rate of Convergence of Switched Least Squares System Identification for Autonomous Markov Jump Linear Systems. IEEE transactions on Automatic Control, 1-8.

\bibitem{Anna2018} Anna Scampicchio, Alberto Giaretta, \& Gianluigi Pillonetto. (2018). Nonlinear Hybrid Systems Identification using Kernel-Based Techniques. In IFAC-PapersOnline, 51(15), 269-274.

\bibitem{Angelo2010} Angelo Alessandri, Marco Baglietto, \& Giorgio Battistelli. (2010). A maximum-likelihood Kalman filter for switching discrete-time linear systems. Automatica, 46, 1870-1876.

\bibitem{Tsurumoto2022} K. Tsurumoto, W. Ohnishi, T. Koseki, N. Strijbosch, \& T. Oomen. (2022). A non-causal approach for suppressing the estimation delay of state observer. 022 American Control Conference (ACC), 3356-3356.

\bibitem{El2023} El Ammari Anis, \& Terzi, Chokri. (2023). Causal Nexus Between Ownership Structure, Dividend Policy and Financial Performance: A Bootstrap Panel Granger non-causality Analysis. Journal of African Business, 24(4), 562-579.


\bibitem{Verhaegen1996} Verhaegen, M. (1996). A subspace model identification solution to the identification of mixed causal, anti-causal LTI systems. SIAM Journal on Matrix Analysis and Applications, 17(2), 332–347.

\bibitem{Fang2024} X. Fang, \& T. Chen. (2024). On kernel design for regularized non-causal system identification. Automatica, 159, 111335.

\bibitem{Blanken2020} Blanken, L., \& Oomen, T. (2020). Kernel-based identification of non-causal systems with application to inverse model control. Automatica, 114.

\bibitem{sta_as1} T. E. Duncan, \& B. Pasik-Duncan. (1990). Adaptive control of continuoustime linear stochastic systems. Math. Control signals systems, 3(1), 45–60.
\bibitem{sta_as2} M. K. S. Faradonbeh, A. Tewari, \& G. Michailidis. (2020). On adaptive linear–quadratic regulators. Automatica, 117, 108982.

\bibitem{Dempster1977} Dempster, Arthur P., Laird, Nan M., \& Rubin, Donald B. (1977). Maximum likelihood from incomplete data via the EM algorithm. Journal of the Royal Statistical Society. Series B. Statistical Methodology, 1–38.




%
%
%
%
%
%

\bibitem{Lai1982} T. L. Lai, \& C. Z. Wei. (1982). Least squares estimates in stochastic regression
models with applications to identification and control of dynamic
systems. Ann. Statist., 10(1), 154–166.

\bibitem{PE2018} P. E. Caines. (2018). Linear stochastic systems. SIAM.
\bibitem{Chen1986} H. F. Chen, \& L. Guo. (1986). Convergence rate of least-squares identification and adaptive control for stochastic systems. Int J Control, 44(5), 1459–1476.

\bibitem{Stout1974} W. F. Stout. (1974). Almost Sure Convergence. Academic Press.

\bibitem{Gibson2005} Stuart Gibson, \& Brett Ninness. (2005). Robust maximum-likelihood estimation of multivariable dynamic
systems. Automatica, 41, 1667-1682.

\bibitem{Kalman} Kalman RE. (1960). A new approach to linear fltering and prediction problems. J Basic Eng, 82(1), 35–45.


\end{thebibliography}

\appendix
\section{Technical proofs}\label{sec:proofs}
\textit{Proof of Lemma~\ref{lemma:1}:} The log likelihood difference between the $\theta$ and $\theta^{k}$ can be expressed as 
\begin{align*}
	\ln \mathbb P_\theta(\bm{y})-\ln \mathbb P_{\theta^k}[\bm{y}]=Q(\theta,\theta^k)-&Q(\theta^k,\theta^k)\\
	&+V(\theta,\theta^k)-V(\theta^k,\theta^k),
\end{align*}
where the difference $V(\theta,\theta^k)-V(\theta^k,\theta^k)$ coincides with the Kullback–Leibler distance that possess an important property, i.e., being non-negative. Therefore, the maximization of $Q(\theta,\theta^k)$ can yield an increase in the log-likelihood function $\ln \mathbb P_\theta(\bm{y})$, namely
\[
	Q(\theta,\theta^{k+1})\geq Q(\theta,\theta^{k})\Rightarrow \ln \mathbb P_{\theta^{k+1}}[\bm{y}]\geq\ln \mathbb P_{\theta^k}[\bm{y}],
\]
thus concluding the proof.
\qedsymbol

\emph{Proof of Lemma~\ref{lemma:boundendness}:} Only the boundedness of $\eta_c(t)$ will be proven in detail, since that of $\eta_a(t)$ and $\eta_m(t)$ can be derived in a similar way.

First, we note that $x_c(t-1)$ can be equivalently expressed  as follows:
\[
	x_c(t-1)=\varphi_1(\bm{s}_c)x_c(1)+\varphi_2(\bm{s}_c)\bm{v}_c(1:t-1),
\]
where $\varphi_1(\bm{s}_c)$ and $\varphi_2(\bm{s}_c)$ are shown in A.1, A.2. $\bm{v}_c(1:t-1)\triangleq[v_c(1),\cdots,v_c(t-1)]$. Both matrices are uniquely determined by the switching sequence $\bm{s}_c$ and system matrices $A_c$. Then, one obtains that:
\newcounter{TempEqCnt} 
\setcounter{TempEqCnt}{\value{equation}} 
\setcounter{equation}{0} 
\begin{figure*}[hb] 
	\hrulefill  
	\begin{eqnarray}			\varphi_1(\bm{s}_c)&=&A_c(s_c(2))+A_c(s_c(3))A_c(s_c(2))+\cdots+A_c(s_c(t-1))A_c(s_{t-2})\cdots A_c(s_c(2))\\
		\varphi_2(\bm{s}_c)&=&\begin{bmatrix}
			1+A_c(s_c(2))+A_c(s_c(3))A_c(s_c(2))+\cdots+A_c(s_c(t-1))\cdots A_c(s_c(2))\\
			1+A_C(s_c(3))+\cdots+A_c(s_c(t-1))\cdots A_c(s_c(3))\\
			\vdots\\
			1+A_c(s_c(t-1))\\
			1
		\end{bmatrix}^\top
	\end{eqnarray}
\end{figure*}
\begin{align*}
	\eta_c(t)&=x_c(t)-A(\hat{s}_c(t))x_c(t-1)\\
	&=(A_c(s_c(t))-A(\hat{s}_c(t))) x_c(t-1)+v_c(t)\\
	&=(A_c(s_c(t))-A(\hat{s}_c(t)))\varphi_1(\bm{s}_c)x_c(1)\\
	&\hspace{3.6cm}+\varphi_3(\bm{s}_c)\bm{v}_c(1:t),
\end{align*}
where $\varphi_3(\bm{s}_c)=[\varphi_2(\bm{s}_c),1]$. Passing to the (squared) norm in the expression above we note that, in view of the fact that the noise $\bm{v}_c$ has a bounded covariance, the last term is bounded too. For what concerns the first term, instead, we have:
\[
	\Vert [A_c(s_c(t))-A(\hat{s}_c(t))]\varphi_1(\bm{s})x_c(1)\Vert^2\leq\lambda_1\Vert x_c(1)\Vert^2,
\]
where 
\begin{align*}
	\lambda_1\triangleq\lambda_{\textrm{max}}(\varphi^\top_1(\bm{s})&(A_c(s_c(t))-A(\hat{s}_c(t)))^\top\\
	&\hspace{1.5cm}(A_c(s_c(t))-A(\hat{s}_c(t)))\varphi_1(\bm{s})),
\end{align*}
which concludes the proof.
\qedsymbol

\emph{Proof of Theorem~\ref{th3}:} In the interest of space, we establish the convergence rate for $\hat{A}_a(l)$ only, since the other bounds on the system matrices can be derived similarly.

In view of Lemma~\ref{lemma1}, sufficient conditions for establishing the convergence rate of $\hat{A}_a(l)$ are (C1) $\lambda_{\textrm{min}}(W^a_{l,T})\rightarrow\infty$, a.s., and (C2) $\log\lambda_{\textrm{max}}(W^a_{l,T})=o(\lambda_{\textrm{min}}(W^a_{l,N}))$, a.s.. We therefore have to show that these two conditions are verified in our case. Then, for what concerns (C1), one has:
\begin{align*}
	x_a(t)x_a(t)^\top&=(\hat{A}_a(l)x_a(t+1)+v_a(t))\\
	&\hspace{2.7cm}(\hat{A}_a(l)x_a(t+1)+v_a(t))^\top\\
	&=\hat{A}_a(l)x_a(t+1)x_a^\top(t+1)\hat{A}^\top_a(l)\\
	&\hspace{1cm}+2v_a(t)x^\top_a(t+1)\hat{A}^\top_a(l)+v_a(t)v_a^\top(t).
\end{align*}
Since $\hat{A}_a(l)x_a(t+1)x_a^\top(t+1)\hat{A}^\top_a(l)$ is a positive semidefinite matrix, by relying on Lemma \ref{lemma3} we can infer that
\begin{align*}
	W^a_{l,T}&=\sum_{t\in\mathbb{T}^a_{l,T}}x_a(t)x^\top_a(t)\\
	&\succcurlyeq\sum_{t\in\mathbb{T}^a_{l,T}}v_a(t)v_a^\top(t)+x_a(T)x^\top_a(T)\\
&+\sum_{t\in\mathbb{T}^a_{l,T}}(\hat{A}_a(l)x_a(t\!+\!1)v^\top_a(t)\!+\!v_a(t)x^\top_a(t\!+\!1)\hat{A}^\top_a(l))\\
	&\succcurlyeq\sum_{t\in\mathbb{T}^a_{l,T}}v_a(t)v_a^\top(t)+o(T).
\end{align*}
Then, we readily obtain:
\begin{align*}
	\lim_{\vert \mathbb{T}^a_{l,T}\vert\rightarrow\infty}\inf&\frac{\sum_{t\in\mathbb{T}^a_{l,T}}x_a(t)x^\top_a(t)}{\vert\mathbb{T}^a_{l,T}\vert}\\
	&\succcurlyeq\lim_{\vert \mathbb{T}^a_{l,T}\vert\rightarrow\infty}\inf\frac{\sum_{t\in\mathbb{T}^a_{l,T}}v_a(t)v_a^\top(t)}{\vert\mathbb{T}^a_{l,T}\vert}\succ0.
\end{align*}
Therefore, we can conclude that $\lambda_{\textrm{min}}(W^a_{l,T})\rightarrow\infty$ a.s..

To prove (C2) we note that:
\begin{align*}
	\lambda_{\textrm{max}}(\textstyle\sum_{t\in\mathbb{T}^a_{l,T}}x_a(t)x^\top_a(t))&\leq \textrm{tr}(\textstyle\sum_{t\in\mathbb{T}^a_{l,T}}x_a(t)x^\top_a(t))\\
	&\leq \sum_{i=1}^{T}\Vert x_a(t)\Vert^2=\mathcal{O}(N),
\end{align*} 
where the last equality follows in view of the stability, in average sense, of the NCS-ASM in \eqref{eq:1}.
Then, one can readily obtain that
\begin{eqnarray}\label{eq6.1}
	\lim_{T\rightarrow\infty}\frac{\log(\lambda_{\textrm{max}}(W^a_{l,T}))}{\lambda_{\textrm{min}}(W^a_{l,T})}\leq\lim_{T\rightarrow\infty}\frac{\log(T)}{\vert\mathbb{T}_{i,T}\vert}=\frac{\log(T)}{\mathcal{O}(T)}=0.\nonumber\\
\end{eqnarray}

We are now able to establish the convergence rate for the covariance matrices. Specifically, we will give the detailed proof for $\hat{\Sigma}_c(j)$ only, since the remaining ones follow similarly.

From the NCA-ASM in \eqref{eq:1}, the true covariance matrix for $v_c$ can be expressed as:
\begin{align*}
	\Sigma_c(j)=\frac{1}{\vert\mathbb{T}^c_{j,T}\vert}\sum_{t\in\mathbb{T}^c_{j,T}}(x_c(t)-&A_c(j)x_c(t-1))\\&(x_c(t)-A_c(j)x_c(t-1))^\top.
\end{align*}
Then, the estimation error can take the following form: 
\begin{align*}
	\hat{\Sigma}_c(j)-\Sigma_c(j)=&\frac{1}{\vert\mathbb{T}^c_{j,T}\vert}\sum_{t\in\mathbb{T}^c_{j,T}}((A_c(j)-\hat{A}_c(j))x_c(t-1))\\
	&\hspace{2cm}((A_c(j)-\hat{A}_c(j))x_c(t-1))^\top.
\end{align*}
Therefore, the convergence rate for $\hat{\Sigma}_c(j)$ reads as:
\begin{align*}
	\Vert\hat{\Sigma}_c(j)-\Sigma_c(j)\Vert_\infty\leq& \frac{\sum_{t\in\mathbb{T}^c_{j,T}}x_c(t-1)x_c(t-1)^\top}{\vert\mathbb{T}^c_{j,T}\vert}\\
	&\Vert (A_c(j)-\hat{A}_c(j))(A_c(j)-\hat{A}_c(j))^\top\Vert_\infty\\
	&\leq\mathcal{O}\left(\frac{\log(T)}{T}\right),
\end{align*}
which completes the proof.
\qedsymbol

\end{document}